\RequirePackage{ifpdf}
\documentclass[12pt,letterpaper]{article}
\pdfoutput=1
\usepackage{jheppub}
\usepackage{epsfig}
\usepackage{enumitem}
\usepackage[latin1]{inputenc}
\usepackage{bbm,amsfonts}
\usepackage{rotating,graphicx}
\usepackage{amssymb,amsmath,amsfonts,mathtools}
\usepackage{fancybox,diagbox}
\usepackage{enumerate}
\usepackage{dsfont}
\usepackage{verbatim}
\usepackage{wrapfig}
\usepackage{slashed}
\usepackage{shuffle}
\usepackage{subcaption}
\usepackage{braket}
\usepackage{pdflscape}
\usepackage{accents}
\usepackage{afterpage}

\author[a]{Marco S. Bianchi}
 
\affiliation[a]{Instituto de Ciencias F\'isicas y Matem\'aticas, Universidad Austral de Chile, Casilla 567, Valdivia, Chile}

\emailAdd{marco.bianchi@uach.cl}  
  
\abstract{We present the three-point function of two spin-two and one scalar twist-two operators in $\mathcal{N}=4$ SYM up to three perturbative orders at weak coupling, obtained via a direct Feynman diagrammatic calculation.}

\title{Two spinning Konishi operators at three loops} 

\def\Tr{\textrm{Tr}}

\numberwithin{equation}{section}

\newlength{\dhatheight}

\begin{document}

\maketitle
\allowdisplaybreaks

\section{Motivation}

In a conformal field theory three-point functions constitute the basic building blocks for correlation functions of operators, which are the fundamental observables. Along with the spectrum of operators, the structure constants of three-point functions completely determine a conformal field theory.
In ${\cal N}=4$ SYM theory structure constants have been the object of integrability \cite{Basso:2015zoa}, fueling the effort of solving the model completely, thanks to the conjectured integrability of the model \cite{Beisert:2010jr}.
In principle, integrability provides a non-perturbative reformulation of the problem, potentially allowing for their determination even at finite coupling \cite{Basso:2022nny}. Three-point functions in ${\cal N}=4$ SYM are therefore meaningful and interesting objects to study.

In this paper we focus on their perturbative determination at weak coupling.
Apart from integrability, more generally the calculation of the structure constants of three-point functions in a conformal field theory can be attacked thanks to the operator product expansion (OPE) \cite{Dolan:2001tt,Dolan:2004iy}, provided the relevant higher-point functions are known. Indeed, the most impressive perturbative results for three-point functions in ${\cal N}=4$ SYM have been derived in this way.

A more traditional approach would be a direct calculation using Feynman diagram expansion. Such a procedure is rather impractical, though. The technical difficulty lies in evaluating complicated three-point integrals. As proposed in \cite{Plefka:2012rd}, a limit can be taken consistently which streamlines the calculation of structure constants, reducing their determination to a two-point problem and taking advantage of powerful techniques developed for propagator integrals. We adopt this strategy here, following previous work \cite{Bianchi:2019jpy}. 
We focus on the same setting as in that paper, namely three-point functions of twist-two operators, endowed with spin. Since the three-point functions involving only one operator with spin have already been computed via the OPE up to five loops \cite{Eden:2012rr,Eden:2016aqo,Goncalves:2016vir,Basso:2017muf,Georgoudis:2017meq}, we focus on correlators with two spinning operators. These have been proven to be computable via the method of \cite{Plefka:2012rd}, mentioned above, up to two loops \cite{Bianchi:2019jpy}. We push here the calculation to three loops, where no explicit result is available yet. We start from the lowest spins, namely three-point functions involving two twist-two Konishi descendants of spin 2.

Three-point functions of multiple twist-two spinning operators have been studied recently \cite{Bercini:2020msp,Bercini:2021jti,Bercini:2022gvs} via integrability and the OPE, developing novel techniques and uncovering interesting properties. Especially, their large spin limit was bootstrapped exactly and was related to null Wilson loops. This provides additional interest in their study.
Finally, the result presented in this note should be calculable independently via the OPE, thanks to the recent results of \cite{Bargheer:2022sfd}, determining higher-point correlation functions of protected operators up to three loops. 

\section{Main result}

We work in $\mathcal{N}=4$ SYM with $SU(N)$ gauge group and coupling constant $g$. The perturbative expansion organizes neatly in terms of the 't Hooft coupling $\lambda =\frac{g^2 N}{16 \pi ^2}$. Up to the order we are working at, no color subleading corrections pop up for the three-point functions we intend to calculate.

In a conformal field theory, as $\mathcal{N}=4$ SYM, the structure of three-point functions is fixed by conformal symmetry \cite{DiFrancesco:1997nk}.
For three-point functions with two operators endowed with spin at positions $x_1$ and $x_3$ and a scalar operator at $x_2$, the problem we focus in this note, the spacetime structure reads \cite{Sotkov:1976xe}
\begin{align}\label{eq:3ptstructure}
&\left\langle \mathcal{O}_{j_1}(x_1) \mathcal{O}_{0}(x_2) \mathcal{O}_{j_2}(x_3) \right\rangle = \!\!\displaystyle
\sum_{l=0}^{\text{min}(j_1,j_2)} C_{j_10j_2}^l\, \frac{\left(Y_{32,1}\cdot z_1\right)^{j_1-l}\,  \left(Y_{12,3}\cdot z_2\right)^{j_2-l}}{|x_{13}|^{2l}} I_{13}^l\nonumber\\& ~~\times
|x_{12}|^{j_1-j_2-\Delta_{12,3}} |x_{23}|^{j_2-j_1-\Delta_{23,1}} |x_{13}|^{j_1+j_2-\Delta_{31,2}}
\end{align}
The conformal invariants $Y$ and $I$ can be written down explicitly
\begin{align}\label{eq:invariants}
& Y_{ij,k}^\mu \equiv x_{ik}^\mu\, |x_{ik}|^{-2} - x_{jk}^\mu\, |x_{jk}|^{-2}\nonumber\\&
I_{ij} = z_{1}\cdot z_{2} - 2\, \left(x_{ij}\cdot z_1\right)\, \left(x_{ij}\cdot z_2\right)\, |x_{ij}|^{-2}\nonumber\\
&x_{ij} \equiv x_i - x_j \qquad,\qquad \Delta_{ij,k} \equiv \Delta_i+\Delta_j-\Delta_k
\end{align}
in terms of a pair of null momenta $z_1$ and $z_2$, whose role will be specified further in the following section. The conformal dimensions $\Delta$ of the operators include their anomalous contribution.
The coefficients $C$ are the structure constants. They are functions of the coupling constant, but they do not depend on space-time coordinates.
Let us separate the structure constant into its classical part and its quantum corrections in the following way
\begin{equation}
C(\lambda,N) = {\cal C}^{(0)} \, {\cal C}(\lambda,N) = {\cal C}^{(0)} \left( 1+ {\cal C}^{(1)} \lambda+ {\cal C}^{(2)}\lambda^2 + {\cal C}^{(3)} \lambda^3 + \mathcal{O}\left(\lambda^4\right)  \right)
\end{equation}
This amounts to normalizing the structure constant by its tree-level value, which offers a more universal result, less dependent on conventions for the operators' definition.
We are interested in determining the quantum corrections up to three perturbative orders.

In the case of three-point functions involving up to one spinning operator, only one structure constant appears. On the contrary, for three-point functions involving two Lorentz tensors of spins $j_1$ and $j_2$, $\min(j_1,j_2)+1$ different tensor structures show up, each multiplying a structure constant, indexed by $l$ in \eqref{eq:3ptstructure}. 
We limit the discussion to spin-2 operators (Konishi descendants in the $sl(2)$ sector of $\mathcal{N}=4$ SYM operators). Hence we aim at computing three structure constants, corresponding to the different allowed polarizations of operators.

This note reports their calculation up to three-loop order at weak coupling, a novel result and the first direct computation of three-loop structure constants, without employing the operator product expansion.
The final result reads\footnote{Version 1 of this submission contains errors in the $\lambda^3$ results for $\mathcal{C}_{202}^{0}$, and $\mathcal{C}_{202}^{2}$. These were caused by a subtle scheme dependence affecting a small set of contributions, which was not treated correctly. Apologies for that.} 
\begin{align}\label{eq:result}
& \mathcal{C}_{202}^{0} = 1-12 \lambda+147 \lambda^2-1830 \lambda^3 + \mathcal{O}\left(\lambda^4\right)\\
& \mathcal{C}_{202}^{1} = 1-6 \lambda+\frac{111}{2}\lambda^2-582 \lambda^3+ \mathcal{O}\left(\lambda^4\right)\\
& \mathcal{C}_{202}^{2} = 1+6 \lambda-87 \lambda^2+1104 \lambda^3+ \mathcal{O}\left(\lambda^4\right)
\end{align}
This result can be potentially used for comparison with a calculation from the OPE expansion of the recently determined five-point functions of protected operators at three loops \cite{Bargheer:2022sfd}. Moreover, a test with a computation based on integrability would also be extremely interesting.

The result presented here is a proof of concept that attacking three-loop structure constants with two unprotected twist-two operators is feasible with the proposed method. After optimization, it should permit the evaluation of more such results, involving operators of greater spins.

In the following sections we describe in detail the method implemented for their determination, spell out a few side products of the computation and point out future directions.

\section{Method}

The method employed for the computation is perturbative expansion by Feynman diagrams.
Direct perturbative evaluation of three-point functions is extremely challenging beyond one-loop level.
Besides, in a conformal field theory it is somehow redundant to keep track of the explicit space-time dependence of individual contributions, since the final result has to obey the general form \eqref{eq:3ptstructure}, dictated by conformal symmetry.

We employ the integration trick proposed in \cite{Plefka:2012rd}, whereby an additional space-time integration over the position of one of the operators, say $x_2$, is performed on both sides of \eqref{eq:3ptstructure}. Specializing on two operators with spin two, we obtain
\begin{align}\label{eq:integrated}
\int d^dx_2\left\langle \mathcal{O}_{2}(x_1) \mathcal{O}_{0}(x_2) \mathcal{O}_{2}(x_3) \right\rangle = &
\displaystyle
\sum_{l=0}^{2} C_{202}^l\, \int d^dx_2
\frac{\left(Y_{32,1}\cdot z_1\right)^{2-l}\,  \left(Y_{12,3}\cdot z_2\right)^{2-l}}{|x_{13}|^{2(l+1+\gamma_2)}|x_{12}|^{2} |x_{23}|^{2}} I_{13}^l
\end{align}
where $\gamma_2$ stands for the anomalous dimension of the spin-2 operators.
On the one side, this operation translates into a soft limit in momentum space for the operator whose position is integrated over. This effectively reduces the three-point function problem to a two-point calculation, which is much easier to attack perturbatively.
The details of the perturbative computation are spelled out in the next section.
On the other side of the equal sign of \eqref{eq:integrated}, the integration can be performed directly by standard integration techniques, leading to
\begin{align}\label{eq:integrated2}
\raisebox{-2.75mm}{\includegraphics[scale=0.3]{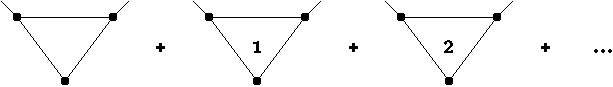}} \, &= \,
\frac{\pi^2}{\epsilon}\left(2 C_{202}^1 -C_{202}^0 - 4 C_{202}^2\right)\frac{(x_{13}\cdot z_1)^2(x_{13}\cdot z_2)^2}{\left|x_{13}\right|^{10}}\nonumber\\&~~~~ + \frac{\pi^2}{\epsilon}\left(4 C_{202}^2-C_{202}^1\right)\frac{(x_{13}\cdot z_1)(x_{13}\cdot z_2)z_{12}}{\left|x_{13}\right|^8}\nonumber\\&~~~~ - \frac{\pi^2}{\epsilon}C_{202}^2\, \frac{z_{12}^2}{\left|x_{13}\right|^6} + \mathcal{O}\left(\epsilon^0\right)
\end{align}
We use the notation $z_1\cdot z_2 = z_{12}$.
As the pictorial equation shows, it is possible to extract the structure constants $C$ by comparing both sides of \eqref{eq:integrated} perturbatively and by powers of $z_{12}$, resulting in a simple linear system of equations.
On the left-hand side, the picture represents the perturbative expansion of three-point functions in the soft limit (no external line for the bottom operator), turning them into two-point problems.
On the right-hand side, the result of the integration in $d=4-2\epsilon$ dimensions appears.

Caution is in order, since various objects involved in the intermediate steps of the computation are potentially divergent.
Dimensional regularization is employed throughout this work, already from the onset of the method \eqref{eq:integrated}.
When expanding the left-hand-side of \eqref{eq:integrated} perturbatively, dimensional regularization is implemented at the level of Feynman diagrams and the resulting Feynman integrals. The $\epsilon\to 0$ limit is performed at the very end of the evaluation.
On the contrary, the $d$-dimensional integration on the right-hand-side of \eqref{eq:integrated} is carried out on a strictly four-dimensional object, that is the general spacetime structure of the three-point function. This means that an $\epsilon\to 0$ limit has been taken before the auxiliary space-time integration.
Strictly speaking, the equality \eqref{eq:integrated} is plagued by an order-of-limits issue which can in principle undermine it completely, unless both sides are finite, which is not the case in the present computation.
Fortunately, it so happens that if the additional integration is performed over the position of the twist-two scalar, then both sides of \eqref{eq:integrated2} present a simple pole in $\epsilon$, but their residues can be compared safely.

There is ample a posteriori evidence that this method works, from lower-loop calculations \cite{Bianchi:2019jpy}. A general criterion for the success of the procedure is presented in \cite{Bianchi:2018zal}. This is based on checking that the order of $\epsilon$ poles does not grow across loop corrections, and it is satisfied in the case at hand.
Violation of this criterion signals that subleading in $\epsilon$ corrections to the right-hand-side of \eqref{eq:integrated} should have been taken into account before the auxiliary integration, spoiling the whole principle which makes the calculation feasible.

It is furthermore conceivable that, even though the above-mentioned condition is satisfied, there could exist evanescent contributions which vanish in dimensional regularization because of the soft limit, but which contribute in exactly four dimensions. 
These would be absent in the left-hand-side of \eqref{eq:integrated} which is strictly in $d=4-2\epsilon$ dimensions, but would contribute to the four-dimensional integrand, before integration.
This mismatch is another product of the order-of-limits issue and an explicit example was uncovered in \cite{Bianchi:2020cfn} in three dimensions.
It is extremely difficult to detect this sort of contributions, especially at high loop order, where the number of diagrams prevents a case-by-case analysis and the process has to be automatized completely. The most compelling test that this potential issue does not manifest in this calculation comes from the fact that three-point functions involving only one spinning operator computed in this way check out against previous independent results \cite{Eden:2012rr}, as shown in more detail below. Since the diagrams contributing to such a three-point function are precisely the same as for the case of three-point functions with two spinning operators, this agreement is a strong hint at the correctness of the calculation.

\section{Implementation and challenges}

\paragraph{Diagrams}
The relevant Feynman diagrams have been produced with QGRAF \cite{Nogueira:1991ex}. Up to three-loop order, there are in principle 5328 diagrams for two-point functions and 28552 diagrams for three-point functions.
The algebra of the Feynman diagrams is performed in momentum space automatically in the computer algebra FORM \cite{Vermaseren:2000nd,Ruijl:2017dtg}, with the help of its \texttt{color} \cite{vanRitbergen:1998pn} library.

A few diagrams produce vanishing results, many could be sifted out as self-energy corrections and some others could be related among each other via symmetries.
We have not yet considered implementing such optimizations to our code, as the calculation presented here is perfectly feasible in a reasonable amount of time.
Still, they will be required for efficiently computing structure constants of operators endowed with greater spin than 2.
We performed the computation for a general covariant gauge, which introduces some additional computational burden, but offers a powerful consistency check.

\paragraph{Regularization}
Dimensional regularization is enforced throughout the calculation.
On the perturbative expansion side, it is required to regulate both ultraviolet divergences stemming from unprotected operators and infrared divergences arising because of the soft limit, apart from intermediate singularities of individual Feynman integrals.

Adopting another regularization scheme is not viable for a number of reasons.
The most pressing is the computational capability offered by dimensional regularization, especially for integration-by-parts (IBP) reduction \cite{Chetyrkin:1981qh,Tkachov:1981wb}.

Dimensional reduction scheme \cite{Siegel:1979wq} is used, whereby metric tensors from vector algebra and integrations are in $d=4-2\epsilon$ dimensions, there are $N_s = 6+2\epsilon$ scalar degrees of freedom, $N_f=4$ fermionic and the gamma matrices algebra is performed in $4$ dimensions.
Such a scheme has been proven to respect supersymmetry in perturbative calculations up to three-loop order in \cite{Velizhanin:2008rw}, producing a vanishing beta function \cite{Avdeev:1981ew}. 

In order to keep track of possible regularization scheme issues, we have initially kept the number of scalars and fermions unfixed, along with the space-time dimensions arising from the algebra of Lorentz vectors and spinors (via gamma matrices) separately.
We observe that the expected vanishing of quantum corrections to two- and three-point functions up to three loops indeed fixes such parameters as dictated by the tenets of dimensional reduction.

As mentioned above, one serious drawback of dimensional regularization is due to potential ambiguities associated with the integration trick. Still, evidence suggests that they do not kick in in the present calculation.

\paragraph{IBP reductions}
Feynman integrals are evaluated efficiently in momentum space, where three-loop corrections to two- and three-point functions entail four-loop integrals.
The computational core and bottleneck lies in the reduction to master integrals.
At four loops 26 propagator master integrals appear in the reduction.
Since the manipulation of the Feynman diagram algebra is already performed using FORM, we chose Forcer as an IBP reducer \cite{Ruijl:2017cxj}, within the same computer algebra. This routine already includes the $\epsilon$ expansion of the relevant master integrals computed in \cite{Lee:2011jt}.
Aside from powers of integral momenta arising from the derivatives in the interaction vertices and propagators, additional powers arise after Fourier transforming the derivatives associated to operators spins.
The latter are contracted with two external auxiliary null momenta $z_1$ and $z_2$. 
In addition to them, the integrals depend on the external momentum $p$, typical of propagator integrals, which is the Fourier transform of the separations $x_{12}$ and $x_{13}$ of two- and three-point functions, respectively.
In order to feed Forcer with such integrals, we previously realize a projection onto scalar integrals depending on the external momentum $p$ only. 
This amounts to inverting linear systems with polynomial entries depending on the dimension $d$.
This is straightforward in the case of one external null momentum, for which a general formula for the coefficients can be determined. For the case of two external null momenta there in no greater conceptual challenge, there are just more coefficients to compute. A practical solution was described in \cite{Bianchi:2022oyz}, though for the low spins we are dealing with it is easy to fix the relevant coefficients analytically (no perturbative solution iterating expansions around $d=4$ is actually required).

\section{The calculation}

A bruteforce evaluation of all required diagrams took a few days on a 32-cores processor with a non-optimal parallelization.
This is a completely feasible computation. Besides, ample margins for optimization exist, which are pointed out below and which we plan to implement, in order to push the computational limits further and be able to derive more results.
In this section we provide more precise details and report some intermediate steps of the calculation, to clarify how it was performed. They are tedious, but could be useful in the unlikely case that somebody tries to replicate the derivation.

\paragraph{Two-point functions.}
In order to renormalize the divergences in three-point functions, and to properly normalize them in a universal manner, an evaluation of the two-point functions of the relevant operators is needed.
Such operators are the twist-two at spin 0 and 2. The twist-two scalar $\Tr(XX)$ is protected and does not mix with other operators. The twist-two operators with spin 2 span a two-dimensional space, since there are two independent derivative combinations $\Tr(D^2XX)$ and $\Tr(DXDX)$. We use complex scalars $X$ in the definition of the operators. In practical calculations we often distinguish the flavors of the two fields, to avoid repeated equivalent contractions. The two choices are equivalent up to overall factors and normalizing by tree level values cancels them away from the final results.
The symbol $D$ denotes the light-cone covariant derivative, explicitly
\begin{equation}
D\, \raisebox{0.5mm}{\scalebox{0.5}{$\bullet$}} = z^\mu\left(\partial_\mu \, \raisebox{0.5mm}{\scalebox{0.5}{$\bullet$}} - i\, g [A_\mu, \, \raisebox{0.5mm}{\scalebox{0.5}{$\bullet$}}]\right)
\end{equation}
contracted with a null auxiliary momentum $z$.
We choose the following basis for the operators
\begin{align}\label{eq:operators}
& O_0 = c^{(0)}\, \Tr(XX)\nonumber\\&
\vec{O}_2 = \left( 
\begin{array}{c}
c^{(2)}_1\, \Tr(D^2XX) + c^{(2)}_2\, \Tr(DXDX)\\
c^{(2)}_3\, \Tr(D^2XX) + c^{(2)}_4\, \Tr(DXDX) 
\end{array}\right)
\end{align}
The conformal operators of definite dimension are linear combinations of \eqref{eq:operators}, obtained via fixing the coefficients $c_i$ in such a way that two-point functions are diagonal.
Such operators are the conformal primary of spin 2 and a descendant of the scalar operator.
Further, we will additionally impose that the operators form a perturbatively orthonormal basis, meaning that their two-point functions evaluate to unity times the correct length scaling and invariant structures dictated by conformal symmetry.
The required order for the proper normalization of operators in dimensional regularization is three powers of the $\epsilon$ parameter beyond the most divergent term at each loop order $l$, which is in general a $l$-order pole in position space.
Practically, this means that we have to evaluate two-point functions up to order $3-l$ at $l$ loops in position space, or equivalently $4-l$ in momentum space (a power of $\epsilon$ arising while Fourier transforming).
At tree level, the coefficients diagonalizing twist-two operators of spin $j$ are expressed in terms of coefficients $C_j^{\frac{d-3}{2}}$ of Gegenbauer polynomials in $d=4-\epsilon$ dimensions
\begin{equation}\label{eq:twist2}
O_j = \sum_{k=0}^j\, a_{jk}^{\frac{d-3}{2}}\, \Tr\left( D^k X D^{j-k} X \right)
\quad\text{where}\quad
\sum_{k=0}^j\, a_{jk}^{\nu}\, x^k y^{j-k} = (x+y)^j\, C_j^{\frac{d-3}{2}}\left( \frac{x-y}{x+y} \right)
\end{equation}
Hence, we define the bare operators \eqref{eq:operators} as
\begin{align}
&c^{(0)} = 1\nonumber\\&
c^{(2)}_1 = 2 (1-\epsilon ) (1-2 \epsilon )\qquad
c^{(2)}_2 = -2 (2-\epsilon ) (1-2 \epsilon )\nonumber\\&
c^{(2)}_3  = c^{(2)}_4  = 2
\end{align}
This choice automatically produces orthonormal two-point functions at tree level.
For correlators where two spinning operators appear, we understand that one is the complex conjugate of the other and they are contracted with different sets of null momenta $z_1$ and $z_2$, satisfying the properties
\begin{equation}
z_1^2= z_2^2 = 0 \qquad\qquad z_1\cdot z_2 = z_{12}
\end{equation}
With the above definitions and absorbing factors of $\left(\frac{e^{\gamma_E}}{(4\pi)}\right)^\epsilon$ as in the $\overline{\text{MS}}$ scheme and additional powers of $|x_{12}|^\epsilon$ as well, the two-point function of the twist-two scalar reads
\begin{equation}
\left\langle O_{0}(x_1) O_{0}(x_2) \right\rangle = \frac{n}{16\pi^4}\, \frac{N^2-1}{\left|x_{12}\right|^{4}}
\end{equation}
where 
\begin{align}
n &= \left(1+\zeta _2 \epsilon ^2+\frac{2}{3}\zeta _3 \epsilon ^3 + \mathcal{O}\left(\epsilon^4\right)\right)
+\left(-24 \zeta _3 \epsilon+\left(24 \zeta _3-36 \zeta _4\right) \epsilon ^2 + \mathcal{O}\left(\epsilon^3\right)\right)\lambda
 \nonumber\\&
+\left(300 \zeta _5 \epsilon + \mathcal{O}\left(\epsilon^2\right) \right)\lambda^2  + \mathcal{O}\left(\lambda^4\right)
\end{align}
As expected, no divergence appears for the scalar operator and moreover the result is subleading in $\epsilon$.
To properly normalize the operator in such a way that subleading in $\epsilon$ corrections are removed from its two-point function we perform the finite renormalization
\begin{equation}
{\cal O}_0 = Z_0\, O_0
\end{equation}
where
\begin{align}
Z_0 &= 1+\left(12 \zeta _3 \epsilon + \left(18 \zeta_4 -12 \zeta _3\right)\epsilon ^2 + \mathcal{O}(\epsilon^3)\right) \lambda \nonumber\\&
+\left(-150 \zeta _5 \epsilon  + \mathcal{O}(\epsilon^2)\right) \lambda ^2
+\mathcal{O}(\epsilon)\mathcal \lambda ^3
+\mathcal{O}(\lambda^4)
\end{align}
The two-point function of twist-two operators of spin 2 depends in general on the tensor structure $I_{12}^2$, defined in \eqref{eq:invariants}. Without loss of generality, the computation can be simplified setting the external null momenta equal $z_1=z_2=z$.
Since there are two twist-two bare operators whose quantum corrections mix, the two-point function is a matrix taking the form
\begin{equation}
\left\langle \vec{O}_{2}(x_1) \vec{O}_{2}(x_2) \right\rangle = \frac{\left(x_{12}\cdot z\right)^4(N^2-1)}{16\pi^4\,\left|x_{12}\right|^{12+2\gamma_2}}
\left(
\begin{array}{cc}
A & B\\
0 & D
\end{array}\right)
\end{equation}
where
\begin{align}
&A = \left(24-172 \epsilon+\left(24 \zeta _2+488\right) \epsilon ^2+\left(-172 \zeta _2+16 \zeta _3-700\right) \epsilon ^3 + \mathcal{O}\left(\epsilon^4\right)\right) \nonumber\\&
+
\left(-\frac{288}{\epsilon } +2112+ \left(-432 \zeta _2-576 \zeta _3-5864\right) \epsilon
\right.\nonumber\\&\left.
+\left(3168 \zeta _2+\frac{17136}{5} \zeta _3-864 \zeta _4+\frac{40668}{5}\right) \epsilon ^2 + \mathcal{O}\left(\epsilon^3\right)\right)\lambda \nonumber\\&
+
\left(\frac{1728}{\epsilon ^2}-\frac{12384}{\epsilon } +  \left(3456 \zeta _2+8640 \zeta _3+31128\right)
\right.\nonumber\\&\left.
+\left(-24768 \zeta _2-\frac{293472}{5} \zeta _3+12960 \zeta _4+7200 \zeta _5-\frac{191696}{5}\right) \epsilon + \mathcal{O}\left(\epsilon^2\right)\right)\lambda^2 \nonumber\\&
+
\left(-\frac{6912}{\epsilon ^3}+\frac{46080}{\epsilon ^2}-\frac{17280 \zeta _2+62208 \zeta _3+93408}{\epsilon}
\right.\nonumber\\&\left.
+\left(115200 \zeta _2+\frac{2155392}{5} \zeta _3-93312 \zeta _4-92160 \zeta _5+\frac{467456}{5}\right) + \mathcal{O}\left(\epsilon\right)\right)\lambda^3  + \mathcal{O}\left(\lambda^4\right)
\end{align}
\begin{align}
&B= \left(-240 \epsilon + 564 \epsilon ^2 + \mathcal{O}\left(\epsilon^3\right)\right)\lambda + \left(720 + 624 \epsilon+\mathcal{O}\left(\epsilon^2\right)\right)\lambda^2 \nonumber\\&
+ \left(-\frac{1440}{\epsilon }-6840+\mathcal{O}\left(\epsilon\right)\right)\lambda^3  + \mathcal{O}\left(\lambda^4\right)
\end{align}
\begin{align}
& D= \left(120-308 \epsilon+\left(120 \zeta _2+284\right) \epsilon ^2+\left(-308 \zeta _2+80 \zeta _3-112\right) \epsilon ^3  + \mathcal{O}\left(\epsilon^4\right)\right) \nonumber\\&
+
\left(-2880 \zeta _3 \epsilon + \left(13968 \zeta _3-4320 \zeta _4\right) \epsilon ^2 + \mathcal{O}\left(\epsilon^3\right)\right)\lambda \nonumber\\&
+
\left(36000 \zeta _5 \epsilon + \mathcal{O}\left(\epsilon^2\right)\right)\lambda^2
+  \mathcal{O}\left(\epsilon^3\right)\, \lambda^3 + \mathcal{O}\left(\lambda^4\right)
\end{align}
The result is divergent and exhibits mixing.
In order to remove divergences and find the eigenfunctions of the dilatation operator, we perform the renormalization
\begin{equation}\label{eq:renormalization}
\vec{\cal O}_2 = Z_{2}\, \vec{O}_2 
\end{equation}
in terms of a $2\times 2$ matrix $Z_2$.
The remaining finite and subleading parts are also included in the definition of the renormalization matrix, in such a way to obtain an orthonormal basis of operators with respect to two-point functions.
The renormalization matrix takes the form
\begin{equation}\label{eq:norm2}
Z_2 = \left(
\begin{array}{cc}
(Z_2)_{11}& (Z_2)_{12}\\
0 &  (Z_2)_{22}
\end{array}\right)
\end{equation}
where
\begin{align}
(Z_2)_{11} &= 1
+\left(\frac{6}{\epsilon}-1+\epsilon \left(3 \zeta _2+12 \zeta _3-7\right) \right. \nonumber\\&\left.~~~~
+\frac{\epsilon^2}{60} \left(-30 \zeta _2+636 \zeta _3+1080 \zeta _4-1457\right)+\mathcal{O}\left(\epsilon^3\right)\right)\lambda \nonumber\\&
+\left(\frac{18}{\epsilon ^2}-\frac{18}{\epsilon }+ \left(18 \zeta _2+36 \zeta _3-41\right)
\right. \nonumber\\&\left.~~~~
+\epsilon\left(-18 \zeta _2+\frac{558 \zeta _3}{5}+54 \zeta _4-150 \zeta _5-\frac{1379}{20}\right)+\mathcal{O}\left(\epsilon^2\right)\right)\lambda ^2 \nonumber\\&
+\left(\frac{36}{\epsilon ^3}-\frac{90}{\epsilon ^2}+\frac{54 \zeta _2-52}{\epsilon }-135 \zeta _2+\frac{3054 \zeta _3}{5}-780 \zeta _5+\frac{702}{5}
+\mathcal{O}(\epsilon)\right)\lambda ^3 \nonumber\\&
+\mathcal{O}(\lambda^4)
\end{align}
\begin{align}
(Z_2)_{12} &= \left(2\epsilon+\frac{13 \epsilon ^2}{30}+\mathcal{O}(\epsilon^3)\right)\lambda \nonumber\\&
+\left(6-20 \epsilon+\mathcal{O}(\epsilon^2)\right)\lambda ^2
+\left(\frac{12}{\epsilon }-58+\mathcal{O}(\epsilon)\right)\lambda ^3
 \nonumber\\&
+\mathcal{O}(\lambda^4)
\end{align}
\begin{align}
(Z_2)_{22} &= 1+
\left(12 \zeta _3 \epsilon+\left(18 \zeta _4-\frac{137 \zeta _3}{5}\right) \epsilon ^2+\mathcal{O}(\epsilon^3)\right)\lambda \nonumber\\&
+\left(-150 \zeta _5 \epsilon +\mathcal{O}(\epsilon^2) \right) \lambda ^2
+\mathcal{O}(\lambda^4)
\end{align}
For the spin-2 operator, extracting the divergent part from the logarithm of the renormalization matrix yields the anomalous dimension
\begin{equation}\label{eq:gamma2}
\gamma_2 = 12 \lambda- 48 \lambda ^2+ 336 \lambda ^3+\mathcal{O}(\lambda^4)
\end{equation}
in agreement with previously computed results \cite{Kotikov:2004er}.
With such a renormalization the two-point functions are diagonalized and normalized to 1 up the required orders in the perturbative expansion and dimensional regularization.

\paragraph{Three-point functions.}
With the bare spin-0 and spin-2 operators in \eqref{eq:operators} we evaluate the following integrated three-point functions: $\int d^dx_2\, \langle {O}_0(x_1) {O}_0(x_2) {O}_0(x_3) \rangle$, $\int d^dx_2\, \langle \vec{O}_2(x_1) {O}_0(x_2) {O}_0(x_3) \rangle$ and $\int d^dx_2\, \langle \vec{O}_2(x_1) {O}_0(x_2) \vec{O}_2(x_3) \rangle$.

The result for $\int d^dx_2\, \langle {O}_0(x_1) {O}_0(x_2) {O}_0(x_3) \rangle$ has finite in $\epsilon$ quantum corrections. After conversion to the structure constant, this yields a tree-level exact result, in agreement with the prediction for protected operators \cite{Intriligator:1999ff}.
For one spinning operator we calculate the vector $\int d^dx_2\, \langle \vec{O}_2(x_1) {O}_0(x_2) {O}_0(x_3) \rangle$ and multiply by the renormalization matrices, in such a way to obtain the renormalized three-point function.
This reads (let us set $|x_{13}|=x_{13}\cdot z_1 = x_{13}\cdot z_2=1$ from now on, to spare some clutter in the expressions)
\begin{align}
&\int d^dx_2\, \langle \vec{\cal O}_2(x_1) {\cal O}_0(x_2) {\cal O}_0(x_3) \rangle =\nonumber\\&~~~~= -\frac{N^2-1}{64\pi^4\epsilon}\left(\begin{array}{c} 
8-48 \lambda+\left(288 \zeta _3+528\right) \lambda ^2+\left(192 \zeta _3-4800 \zeta _5-6048\right) \lambda ^3\\
8\end{array}\right)\nonumber\\&~~~~+\mathcal{O}(\epsilon^0)+O\left(\lambda ^4\right)
\end{align}
Finally, by focussing on the first component of the vector and comparing to \eqref{eq:integrated2}, we extract the structure constant already known from previous computations \cite{Eden:2012rr} and reported below.

Finally, we consider the integrated correlator $\int d^dx_2\, \langle \vec{O}_2(x_1) {O}_0(x_2) \vec{O}_2(x_3) \rangle$. Its expression is definitely too long to be displayed here.
After multiplication by the normalization matrices and selecting the upper-left component, corresponding to the three-point function involving conformal primary operators, we find
\begin{align}
\int d^dx_2\, \langle \vec{\cal O}_2(x_1) {\cal O}_0(x_2) \vec{\cal O}_2(x_3) \rangle_{11} =& -\frac{N^2-1}{64\pi^4\epsilon}\left(384-192 z_{12}+16 z_{12}^2\right)\nonumber\\&-\frac{N^2-1}{64\pi^4\epsilon} \left(-1920+384 z_{12}+96 z_{12}^2\right)\lambda\nonumber\\&
-\frac{N^2-1}{64\pi^4\epsilon} \left(18048-1536 z_{12}-1392 z_{12}^2\right)\lambda ^2 \nonumber\\&- \frac{N^2-1}{64\pi^4\epsilon}\left(-195456 + 3840 z_{12} + 17664 z_{12}^2\right)\lambda ^3\nonumber\\&+\mathcal{O}(\epsilon^0)+
\mathcal{O}(\lambda^4)
\end{align}
Comparing the general form of the integrated three-point function \eqref{eq:integrated2} to the above result, we extract the three-point functions for the spin-2 operators
\begin{align}
& \mathcal{C}_{202}^{0} = 1-12 \lambda+147 \lambda^2-1830 \lambda^3 + \mathcal{O}\left(\lambda^4\right)\nonumber\\
& \mathcal{C}_{202}^{1} = 1-6 \lambda+\frac{111}{2}\lambda^2-582 \lambda^3+ \mathcal{O}\left(\lambda^4\right)\nonumber\\
& \mathcal{C}_{202}^{2} = 1+6 \lambda-87 \lambda^2+1104 \lambda^3+ \mathcal{O}\left(\lambda^4\right)\nonumber
\end{align}
which is the main result of this note, reported above.

\section{Tests and additional results}

The complexity of the calculation and regularization subtleties demand extensive consistency checks, which are reported here. They are all successful.
\paragraph{Gauge invariance.} We performed the calculation of both two-point and three-point functions with generic covariant R$_\xi$ gauge. The dependence on the gauge parameter correctly drops out in the expression for the correlators, which have to be gauge invariant. This test also probes the correct definition of the operators \eqref{eq:operators}, which have to be expanded perturbatively in the gauge fields appearing in the covariant derivatives and give rise to additional diagrams.
\paragraph{Protected correlators.} Both two- and three-point functions of protected operators are expected not to develop quantum corrections. We checked that this is the case up to three-loop order. This is a nontrivial test and the vanishing of such quantum corrections depends crucially on a sensible choice of the regularization scheme, which is dimensional reduction. In particular, the vanishing only occurs if the number of bosonic degrees of freedom precisely matches the fermionic ones, which is the purpose of that scheme and requires a non-integer number of scalars in loops.
\paragraph{Renormalization.} Two-point functions for the spin-2 operator develop singularities which require renormalization of the bare operators. The anomalous dimension of the operator extracted in this way \eqref{eq:gamma2} coincides with previously derived results \cite{Kotikov:2004er}. The bare three-point functions integrated over the position of the protected operator renormalize correctly when multiplied by the renormalization matrices \eqref{eq:norm2}. The leftover presents a simple pole in dimensional regularization, which is expected to arise when integrating a finite result over $x_2$ \eqref{eq:integrated2}.

\paragraph{Three-point function for one spinning operator.} The case for only one spinning operator has been computed at three loops and for any spin in \cite{Eden:2012rr}, via an OPE expansion. We performed the calculation with our method, finding agreement with that result 
\begin{equation}
{\cal C}_{200} = 1-6 \lambda+ (36 \zeta_3+66)\lambda^2+(24 \zeta_3-600 \zeta_5-756)\lambda^3 + \mathcal{O}\left(\lambda^4\right)
\end{equation}
which is a promising indication of correctness and an independent check, albeit rather late and unnecessary of \cite{Eden:2012rr}.

\paragraph{Rationality.} Albeit ubiquitous in intermediate steps of the calculation, transcendental zeta values drop out of the final result for structure constants. This pattern is expected for this particular three-point function with two equal spins, as observed at lower loop order \cite{Bianchi:2019jpy}. There is also an explanation based on integrability, which is however beyond our full control and understanding. Anyway, the cancellation of zeta values constitutes a successful consistency check.
The dependence of structure constants on transcendental values is expected to manifest itself explicitly in the case of three-point functions with two operators of different spin.

\paragraph{Aesthetics.} This check is surely unscientific, however, when in individual Feynman diagrams rational numbers with up to 12 digits in numerators and denominators are encountered and they finally sum up to integers or half integers in the final result, the feeling is that the calculation is correct.

\section{Future directions}

We point out some future directions.
Additional structure constants for operators of greater spins are within the computational reach of the algorithm presented in this note.
For their determination in a reasonable time span, a few optimizations are in order.
Some concern the expansion into diagrams.
Self-energies can be filtered out and precomputed, especially three-loop corrections to the scalar propagators, which constitute a considerable portion of the total diagrams.
Diagrams giving rise to a posteriori vanishing or subleading contributions in dimensional regularization may be eliminated, from the onset, instead of computing them with additional momenta.
Diagrams with similar integral topologies may be summed before insertions of the extra momentum factors emerging from spinning operators, minimizing the number of repeated tasks.
Some contributions can be related to each other via reflection symmetries of the integrals.  
Once it is no longer needed for checks, explicit gauge dependence could be dropped in favor of the more economic Feynman gauge, which lightens the algebra of diagrams with several gauge corrections.

The following optimizations in the IBP reductions are also possible.
In order to be on the safe side, in the present calculation the orders of the truncation of the expansion in the dimensional regularization parameter have been overestimated, especially in the expansion of rational factors emerging in the intermediate steps of IBP reductions. There are margins for optimization on this side of the computation, after an analysis of the thresholds for a consistent calculation.

The computation has already been heavily parallelized via bash scripting, treating different diagrams as independent sequential FORM tasks.
For the extension to three-point functions with greater spins, it may be necessary to parallelize into less disk consuming processes, in order not to hit disk capacity limits. A more efficient task distribution could be designed, optimizing the parallelization, or we could switch entirely to the parallel versions of FORM \cite{Tentyukov:2007mu}.
We plan to implement such optimizations in the future, and deliver more results.

Apart from improving the capability of our calculation, it would be rewarding to compare the results to an OPE derivation from the higher-point correlation functions of \cite{Bargheer:2022sfd} and from an integrability standpoint.

\acknowledgments

This work was supported by Fondo Nacional de Desarrollo Cient\'ifico y Tecnol\'ogico, FONDECYT 1220240. We also thank Beca Santander de Movilidad Internacional, for funding a research visit to Niels Bohr Institute and Icelandic University, where most of this work was performed. We thank these institutions and especially our hosts Charlotte Kristjansen and Valentina Giangreco Marotta Puletti for hospitality.

\vfill
\newpage

\bibliographystyle{JHEP}

\bibliography{biblio2}

\end{document}